\documentstyle[12pt]{article}

\title{Reformulation of QCD in the language of general relativity}

\author{F. A. Lunev \thanks{electronic address:
lunev@hep.phys.msu.su } \\ {\em Physical Department, Moscow State
University,} \\ {\em Moscow, 119899, Russia} }

\date{ \ \ \  }

\begin{document}

\maketitle

\begin{abstract}

It is shown that there exists such collection of variables that the
standard QCD Lagrangian can be represented as the sum of usual
Palatini Lagrangian for Einstein general relativity and the
Lagrangian of matter and some other fields where the tetrad fields
and the metric are constructed from initial $SU(3)$ Yang - Mills
fields.

\end{abstract}

\section{Introduction}

Unified description of all interactions is one of the main goals of
the modern physics. Partial unification, namely unification of
electromagnetic and weak interactions , is achieved in Salam -
Weinberg theory and its numerous modifications. More or less
satisfactory unification of electromagnetic, weak and strong
interactions is achieved in grand unified theories based on various
"large" gauge groups ($SU(5), \ SO(10)$, etc.) But the satisfactory
unified description of electromagnetic, weak, strong, and gravity
interactions is still open problem.

The origin of the difficulties is clear. Whereas all realistic
theories of strong, weak, and electromagnetic interactions are based
on Yang - Mills (YM) action

\begin{equation} S_{YM} = \int \mbox{tr} (dA + A \wedge A) \wedge
*(dA + A \wedge A), \label 1 \end{equation}

\noindent the general relativity is based on Einstein - Hilbert
action

\begin{equation} S_{EH} = \int dx \, \sqrt g R \label 2
\end{equation}

\noindent or, in Palatini formalism, on the action \footnote{We omit
insufficient overall factors before actions (\ref 1)-(\ref 3)}

\begin{equation} S_P = \int \, e^a \wedge e^b \wedge (d \, \Gamma +
\Gamma \wedge \Gamma )^{cd} \varepsilon _{abcd} \label 3
\end{equation}

Obviously, that the mathematical structure of the action (1) and the
actions (2) or (3) is very different. So the origination of the
theory, that reduces to (\ref 1) and (\ref 2) (or (\ref 3)) in
certain limiting cases is a very hard problem.

The most direct way to construct unified theory of all interactions
is, of course, to replace the action (\ref 1) by some gravity-like
action, or, vice-a-versa, to replace the action (\ref 2) or (\ref 3)
by another one, that is more similar to  (\ref 1).

The first possibility is realized, for instance, in tensor dominance
(or strong gravity) model \cite{ISS} (see also \cite{M1} for review
and further references.) The Lagrangian of this model is very
similar to gravitational one, but till now relation of this model
and realistic physical models based on YM action is unclear.

The second possibility is realized, for example, in Poincare gauge
theories of gravitation (see \cite{HNH,M1} for review), or in
$SL(6,C)$ gauge theory of Salam, Isham, and Strathdee \cite{SIS} and
their modifications \cite{M1,M2,HD}. But physical meaning of all
above mentioned theories is not quite clear because the
corresponding actions are unlike the action of Standard Model and it
is not obvious that the latter can be considered as some limiting
case of the former.

There exist also many other approaches to unification of gravity and
 YM gauge theories based on different modifications of actions (\ref
1)-(\ref 3).  But, to author's knowledge, all theories proposed are
rather far from real physics.

But there exist the third way to unification of general relativity
and YM theories. Namely, one can try to find such variables that
standard Einstein - Hilbert or Palatini actions written in these
variables are transformed in standard YM action (plus, may be, the
action with some supplementary fields), or, vice-a-versa, one can
try to transform by change of variables the usual YM action in
Einstein - Hilbert or Palatini ones.

During the last twenty years, and especially during last five years,
the great progress was achieved in both directions.

First of all, author would like to mention the Ne'eman - Sijacki
''chromogravity'' approach to QCD developed in papers \cite{NS}.
Ne'eman and Sijacki showed that there exists the mechanism of
appearance of gravity-like forces in infrared limit of QCD. Some
speculations in spirit of Ne'eman - Sijacki approach were also given
in recent paper of Kuchiev \cite{K}.

But in present paper we will follow another approach, namely the
approach, proposed in the author's paper \cite{L1}.

Let us consider,  first, YM theory and general relativity in three
dimensional space-time\footnote{Interesting approach to unification
of YM theory and general relativity in three dimensions was proposed
by Peldan \cite{Pel}. However, it isn't clear, how to generalize
this approach on 4D case.}.  In this case YM action in the first
order formalism can be written in the form

\begin{equation} S_{YM}= \int \mbox{tr} *F \wedge (dA + A \wedge A)
+ \lambda ^2 \int \mbox{tr} *F \wedge F, \label 4 \end{equation}

\noindent whereas Palatini action for gravity is

\begin{equation} S_{3D} = \int e^a \wedge (d \, \Gamma + \Gamma
\wedge \Gamma )^{bc} \varepsilon _{abc} \label 5 \end{equation}

\noindent In the formula (\ref 4) $\lambda $ means coupling constant
\footnote{We reserve more usual notations $e$ or $g$ for
determinants of the tetrad and the metric respectively.}, and $*$ is
the Hodge operator with respect to the space - time metric $g_{mn}$.
For simplicity, below in this section we will consider the case of
Euclidean space - time.

For $SU(2)$ gauge group, forms $F$ and $A$ valued in the space of
anti-symmetric $3 \times 3$ matrices and so we can write

\begin{equation} *F^{ab} = - \varepsilon ^{abc} *F^c, \label 6
\end{equation}

\begin{equation} S^{YM} = \int *F^a \wedge (dA + A \wedge A)^{bc}
\varepsilon _{abc} - \lambda ^2 \int *F^a \wedge F^a. \label 7
\end{equation}

In three dimensions $*F^a$ are 1-forms and so the first term in
(\ref 7) coincides with three dimensional Palatini actions (\ref 5)
up to notations! This fact allows to formulate 3D YM theory in
general relativity-like form with the tensor

\begin{equation} G_{mn} =(*F ^a)_m (*F ^a)_n  \label 8
\end{equation}

\noindent as the new space - time metric. In particular, usual YM
equations appear to be equivalent to Einstein ones with simple rhs.

Above mentioned results concerning relations between 3D gravity and
3D YM theory were obtained, first, in the author's paper
\footnote{In some aspects, similar results were obtained by Halpern
\cite{Hal} in his investigations of self-dual sector of 4D YM
theory. But Halpern's ''metric'' constracted from YM fields is not
rank two space-time tensor and so it can not be considered as analog
of the metric in general relativity.} \cite{L1}. Independently,
analogous results were obtained also in the work \cite{J} in the
context of (3+1) dimensional $SU(2)$ YM theory in the gauge $A^a_0
=0$. However, in the latter approach YM induced gravity lives only
on the three dimensional hyperplanes $x^0 = const$ and so this
approach is essentially non-covariant.

Further three dimensional space-time geometry discovered in works
\cite{L1,J} was investigated in papers
\cite{L2,L3,HOM,BFH,HJ,RS,FHJL}.  In particular, in paper \cite{L2}
solutions of Euclidean 3D YM equations with singularity on the
sphere were discovered. These solutions can be also interpreted as
stationary solutions of 4D YM equations in the gauge $A_0 = 0$ and
can be considered as analog of Schwartzchild solution in general
relativity. Analogous Schwartzchild-like and Kerr-like solutions of
Yang - Mills - Higgs equations were recently discovered by Singlton
\cite{S}.

It was shown that quantum particle moving in such YM field (that is
considered as external one) inside this sphere can not leave it. So,
may be, such solutions can be used for elaborating of black hole or
microuniverse (see \cite{SS}) mechanism of confinement.

We see that in three dimensional world gravity does live inside YM
theory. It is easy to understand the origin of such YM induced
gravity. Indeed, the usual gravity is described by the triad of
covectors  $e^a$ defined in each point of the space-time up to
$SO(3)$ rotation, and $SO(3)$ connection $\Gamma $ that defines the
parallel transport of tensors in the space-time. All these objects
appear naturally in YM theory -- 1-forms $*F ^a$ play the role of
the triad and $SU(2)$ YM connection $A$ plays the role of the
space-time connection $\Gamma $.

But how to generalize this construction for realistic four
dimensional case? Direct generalization is not possible, because,
first, $*F $ in four dimensions are 2-forms (rather then 1-forms as
in 3D case), and, the second, the structure of 4D Palatini action
(\ref 3) differs from one of 3D action (\ref 5).  Nevertheless, such
generalization exists. Moreover, this problem was partially solved,
in fact, almost twenty years ago in Plebanski's work \cite{P}. But
 Plebanski obtained his results in absolutely different context (he
investigated complex structures in general relativity).  May be, due
to this reason his results haven't been used in investigations of YM
induced gravity till now\footnote{
After the completion of this paper author learned about very recent
works by Robinson \cite{Rob} in which the analogy between general
relativity and YM theory in Plebanski approach is considerably
clarified. The relations of results of \cite{Rob} and ones obtained
in the present paper need further investigations.}.

Let us rewrite the Palatini action (\ref 3) in spinor notations
\footnote{We use the usual isomorphism between the spaces of $O(4)$
vectors and $SU(2) \times SU(2)$ spinors. Sign conventions,
normalization factors, etc. are describe in the section 2 below.
Further, we omit the part of Palatini action that contains the
fields $\Gamma _{A'B'}$ because the full action is equivalent to the
chiral action (\ref 9) . (See, for instance, Refs. \cite{JS,Sam}, in
which it was shown that the using of chiral action (\ref 9) is very
natural, in particular, in Ashtecar formalism.)} :

\begin{equation}S _P = \int e^A_{\ \, C'} \wedge e^{BC'} \wedge (d
\, \Gamma _{AB} + \Gamma _{AC} \wedge \Gamma ^C_{\ \, B }) \label 9
\end{equation}

One can note that 1-forms $e^{AA'}$ enter in action only in the
combination

\begin{equation}\Sigma ^{AB} = e^A_{\ \, C'} \wedge e^{BC'} \label a
\end{equation}

\noindent So the Palatini action (\ref 9) can be represented in the
form

\begin{equation}S _P = \int \Sigma ^{AB} \wedge (d \, \Gamma _{AB} +
\Gamma _{AC} \wedge \Gamma ^C_{\ \, B }) \label b \end{equation}

Of course, the quantities $\Sigma ^{AB} $ in (\ref b) cannot be
considered as the independent dynamical variables. Indeed, due to
(\ref a) the 2-forms $\Sigma ^{AB}$ satisfy the condition

\begin{equation}\Sigma ^{(AB} \wedge \Sigma ^{CD)} =0 \label c
\end{equation}

\noindent Further, Plebanski showed that if the conditions (\ref c)
are satisfied and

\begin{equation}\Sigma ^{AB} \wedge \Sigma _{AB} \not = 0, \label d
\end{equation}

\noindent then $\Sigma ^{AB}$ can be represented in the form (\ref
a) with non-degenerate tetrad $e^{AA'}$.

\noindent So the Palatini action (\ref 9) is equivalent to Plebanski
action

\begin{equation}S_{Pl} = \int \Sigma ^{AB} \wedge (d \, \Gamma _{AB}
+ \Gamma _{AC} \wedge \Gamma ^C_{\ \, B}) + \int \phi _{ABCD} \Sigma
^{AB} \wedge \Sigma ^{CD} \label e \end{equation}

The second term in (\ref e) with totally symmetric Lagrange
multipliers $\phi _{ABCD}$ is introudced to take into account the
condition (\ref c).

In the action (\ref e) fields $\Sigma ^{AB}, \, \Gamma _{AB} $ and
$\phi _{ABCD}$ are independent dynamical variables. The first term
in (\ref e) coincides with the first term in the first order $SU(2)$
YM action

\begin{equation}S_{YM} =\int F^{BC} \wedge (dA_{BC} + A_{BD} \wedge
A^D_{\ \, C}) + \lambda ^2 \int F^{BC} \wedge *F _{BC} \label f
\end{equation}

\noindent up to notations. But it doesn't mean that the gravity
lives inside $SU(2)$ YM theory as in 3D case, because the analog of
the second term in (\ref e) is absent in (\ref f) and so we have no
analog of (\ref c) in $SU(2)$ YM theory. But without the condition
(\ref c) we cannot reconstruct the tetrad $e^{AA'}$.

Let us consider, however, the theory with more large gauge group $G
\supset SU(2) $. One can choose among $N = \dim G$ 2-forms $F$ three
forms $F^{AB}=F^{(AB)}$ that are transformed as rank two symmetric
spinor under gauge transformations from certain $SU(2)$ subgroup of
the group $G$. Then the action (\ref f) will be a piece of the total
YM action. Further, if $\dim G \ge 8$, then, in general, we can
impose, using other gauge degrees of freedom, five $SU(2)$ invariant
gauge conditions

\begin{equation}F^{(AB} \wedge F^{CD)} = 0 \label g \end{equation}

Conditions (\ref g), that we will call "the Plebanski gauge",
coincide with Plebanski conditions (\ref c) up to notations whereas
the first term in YM action (\ref f) coincides, up to notations,
with the first term in Plebanski action (\ref e). {\em So we can
conclude that gravity lives inside YM theory if the dimension of the
gauge group is more or equal to eight.} Indeed, due to Plebanski
theorem we can reconstruct the tetrad $e^{AA'}$ and the
corresponding metric:

\begin{equation}F^{AB} = e^A_{\ \, A'} \wedge e^{BA'} \label h
\end{equation}

\begin{equation}G_{mn} = e^{AA'}_{\quad \, m}e_{AA'n} \label i
\end{equation}

After substituting (\ref h) in the first term of the action (\ref f)
we obtain the usual Palatini action for gravity (\ref 9).

{\em The main idea of the present work is to use of the gauge (\ref
g) to reformulate the YM theory in general relativity-like form.}
Below we will show that the gauge (\ref g) really exists for the
gauge group $SU(3)$ and the corresponding gauge theory, the Quantum
Chromodynamics, can be formulated in the close analogy with general
relativity. But before author would like to give some additional
notes concerning 2-forms formalism in general relativity.

Plebanski results allow to use three 2-forms $\Sigma ^{AB}$ instead
of metric. In Plebanski's approach these forms satisfy the
constraints (\ref c) that play the crucial role in Plenanski's
formalism. But later it was shown that this conditions aren't
necessary. Namely, it appears that, in generic case, any three
2-forms define unique, up to conformal factor, the metric, with
respect to which they are (anti)-self-dual. These statement is known
now as Urbantke theorem (see \cite{U}. Another proof and some
refinements were given in \cite{H} ). In particular, in generic case
any collection of three 2-forms defined up to $SL(3)$ transformation
naturally determine the unique metric.  Moreover, later 't Hooft
showed \cite{'tH} that any triple of two forms (with some
non-degeneracy condition) naturally defines not only the metric but
also certain $SL(3)$ connection and so it is possible to reformulate
the general relativity in terms of triples of 2-forms.  Similar
formalism with $GL(3)$ connection instead of $SL(3)$ ones was
proposed also in recent paper \cite{B}.

However, the Lagrangian of 't Hooft and its modifications are
reduced to Plebanski's Lagrangian (\ref e) by imposing of gauge
conditions that are exactly coincide with (\ref c). On the other
hand, t' Hooft Lagrangian, without the imposing of conditions (\ref
c), is quadrilinear and so is not similar to YM one. By these
reasons in the present paper we use the old Plebanski formalism
rather then its further generalizations.

Clear relations between gravity and YM theory also appear in
Ashtecar formalism \footnote{An attempt to develop the formalism for
unified description of YM and gravity fields in the spirit of
Ashtecar phase space approach was done in works \cite{CP}. However,
this theory gives the conventional YM theory only in the lowest
order in the fields and so it is hard to make consistent its
predictions with ones of the Standard Model.} .  Originally
discovered \cite{A}, it was very unlike Plebanski approach. But
later it was shown \cite{CDJM} that Ashtecar formalism can be
reproduced by (3+1) decomposition of Plebanski Lagrangian.

The paper is organized as follows. In section 2 we describe our
notations. In sections 3 and 4 we formulate QCD in general
relativity-like form at classical amd quantum levels respectively.
In section 5 we discuss obtained results.

\section{Notations.}

Indexes $a,b,c,d$ are frame ones and run over the set $\{ 0,1,2,3
\}.$ Indexes $m,n,p,q$ are world ones and run over the same set.
Upper case Latin indexes $A,B,C, ... $ are $SU(2)$ spinor ones and
run over the set $\{ 0,1 \}.$ Greek indexes $\alpha , \beta , \gamma
$ runs over the set $\{ 1, 2, 3\}$

\subsection{$SU(2)$ spinors and $O(3)$ vectors.}

Lowering and raising of $SU(2)$ spinor indexes are performed by
anti-symmetric spinors $\varepsilon _{AB}$, $\varepsilon ^{AB}$,
$\varepsilon _{01} = \varepsilon ^{01} = +1 $,

\begin{equation}\varphi _A = \varphi ^B \varepsilon _{BA}, \ \
\varphi ^B = \varepsilon ^{BA} \varphi _A \end{equation}

\noindent Hermitian conjugation of $SU(2)$ spinors are defines as

\begin{equation}(\varphi ^{\dag} )^{AB...} = \bar{\varphi}^{CD...}
\varepsilon ^{CA} \varepsilon ^{DB}... \end{equation}

\noindent Here the bar denotes complex conjugation. The spaces of
symmetric second rank $SU(2)$ spinors and $O(3)$ vectors are
isomorphic. The isomorphism is established by the formula

\begin{equation}S^{\alpha } \longleftrightarrow S^{AB} = -
\frac{i}{\sqrt 2} S^{\alpha } \sigma _{\alpha }^{\ AB} \label j
\end{equation}

\noindent where $\sigma _{\alpha \ \, B}^{\ A}$ are Pauli matrices.
Real vectors correspond to Hermitian spinors,

$$\varepsilon ^{\alpha\beta\gamma}U^{\alpha}V^{\beta}W^{\gamma} =
\sqrt 2 U^{AB}V_{BC}W_A^{\ \, C} $$ \begin{equation}S^{\alpha }
 S^{\alpha } = S^{AB} S_{AB}.  \label {j1} \end{equation}

Below we will use the convention (\ref j) with one exception: if
$\Gamma ^{\alpha }$ are components of some $O(3)$ connection and

\begin{equation}R^{\alpha } = d \, \Gamma ^{\alpha } + \frac 1 2
\varepsilon ^{\alpha \beta \gamma } \Gamma ^{\beta } \wedge \Gamma
^{\gamma } \end{equation}

\noindent are components of the corresponding curvature form, then

$$\Gamma ^{AB} = \frac{1}{2i} \sigma _{\alpha }^{\ AB} \Gamma
^{\alpha } $$

\begin{equation}R ^{AB} = \frac{1}{2i} \sigma _{\alpha }^{\ AB} R
^{\alpha } \label k \end{equation}

\noindent Using (\ref k), one can prove that

\begin{equation}R^{AB} = d \, \Gamma ^{AB} + \Gamma ^A_{\ \, C}
\wedge \Gamma ^{CB} \label {kk} \end{equation}

\subsection{$SU(2) \times SU(2) $ spinors and $O(4)$ vectors.}

$O(4)$ frame vector indexes are lowered and raised by the tensor
$\delta ^{ab}$.  The spaces of rank (1,1) $SU(2) \times SU(2) $
spinors and $O(4)$ vectors are isomorphic. The isomorphism is
established by the formula

\begin{equation}S_{AA'} \longleftrightarrow S_a =g_a^{\ AA'} S_{AA'}
, \end{equation}

\noindent where $g_a^{\ AA'}$ are Euclidean Infeld - van der Vaerden
symbols for flat space:

\begin{equation}(g_a^{\ AA'})= \left( \frac{1}{\sqrt 2} \delta^A_{\
\, A'}, \, \frac{i}{\sqrt 2} \sigma _{\alpha \ \, A'}^{\ A} \right)
\end{equation}

Real vectors correspond to Hermitian spinors (Hermitian conjugation
of the latter is defined in the previous subsection),

\begin{equation}S_a S_a = S_{AA'} S^{AA'} \label {k1}\end{equation}

\subsection{$O(4)$ (anti)-self-dual tensors and $O(3)$ vectors}

For frame $O(4)$ tensors Hodge operator is defined as usual:

\begin{equation} *M_{ab} = \frac 1 2 \varepsilon _{abcd} M_{cd}
\end{equation}

The spaces of the (anti)-self-dual tensors and $O(3)$ vectors are
isomorphic. The isomorphism is estublished by the formula

\begin{equation} {}^{\pm }M^{\alpha } = {}^{\pm }\eta ^{\alpha }_{\
ab} {}^{\pm }M_{ab} \end{equation}

\noindent where ${}^{\pm }\eta ^{\alpha }_{\ ab}$ are 't Hooft
symbols:

\begin{equation} {}^{+}\eta ^{\alpha }_{\ ab} = - \frac{1}{2i} \bar
\sigma ^{\ A'}_{\alpha \ \, B'} g_{aAA'} g_b^{\ AB'} \end{equation}

\begin{equation} {}^{-}\eta ^{\alpha }_{\ ab} =  \frac{1}{2i} \sigma
^{\ A}_{\alpha \, \ B} g_{aAA'} g_b^{\ BA'} \end{equation}

't Hooft symbols satisfy the following equations:

\begin{equation} {}^{\pm }\eta ^{\alpha }_{\ ac} {}^{\pm }\eta
^{\beta }_{\ cb} = - \frac 1 4 \delta ^{\alpha \beta } \delta _{ab}
+ \frac 1 2 \varepsilon ^{\alpha \beta \gamma } {}^{\pm }\eta
^{\gamma }_{\ ab} \label {k2} \end{equation}

\begin{equation} {}^{\pm }\eta ^{\gamma }_{\ ab} {}^{\pm }\eta
^{\gamma }_{\ cd} = \frac 1 4 (\delta _{ac} \delta _{bd} - \delta
_{ad} \delta _{bc} \pm \varepsilon _{abcd} ) \label {k3}
\end{equation}

Formulas (\ref{j1}), (\ref{k1}), (\ref{k2}), and (\ref{k3}) allow to
 translate easily any formula from spinor to vector language and
vice-a-versa.

\section{Plebanski gauge in $SU(3)$ Yang-Mills theory}

\subsection{Plebanski theorem for real 2-forms}

Plebanski showed that three complex 2-forms $\Sigma ^{AB}$,
satisfying the conditions (\ref c ) and (\ref d ), can be
represented in the form (\ref a). The "real" variant of this theorem
can be formulated in the following way:

\begin{quote} {\em Let $S^{\alpha }$ be three real 2-forms obeying
the conditions}

\begin{equation} S^{\alpha } \wedge S^{\beta } = \frac 1 3 \delta
^{\alpha \beta }  S^{\gamma } \wedge S^{\gamma }   \label G
\end{equation}

\begin{equation} S^{\gamma } \wedge S^{\gamma } \not = 0 \label H
\end{equation}

\noindent {\em Let  $G_{mn}$ be the Urbantke metric} \footnote{Our
definition of the metric (\ref I) differs from the original
definition of Urbantke \cite{U} by  insufficient factor.}

\begin{equation} G_{mn} = - \frac 4 3 \left[ S^{\delta}_{\ tu}
S^{\delta}_{\ vw} \varepsilon ^{tuvw} \right]^{-1} \varepsilon
_{\alpha \beta \gamma } \varepsilon ^{\, pqrs} S^{\alpha }_{\ mp}
S^{\beta }_{\ qr} S^{\gamma }_{\ sn} \label I \end{equation}

{\em Then $G_{mn}$ has definite signature, $(++++)$ or $(- - - -)$,
and $S^{\alpha }$ can be represented, respectively, as}

\begin{equation} S^{\alpha } = \pm {}^{-} \eta ^{\alpha }_{\ ab} e^a
\wedge e^b . \label J \end{equation}

\end{quote}

One notes, that the equations (\ref G) and (\ref H) are nothing but
reformulation of eqs. (\ref c) and (\ref d) in vector language. The
spinor analog of (\ref J) is

\begin{equation} S^{AB} = \pm \frac 1 2 e^{AC'} \wedge e^B_{\ C'}
\label K \end{equation}

\noindent where

\begin{equation} ( S^{\dag} )^{AB} = S^{AB}, \ \ ( e^{\dag} ) ^{AA'}
= e^{AA'} \end{equation}

\noindent The eq. (\ref K) is the analog of (\ref a).

Let us prove the theorem formulated above. Let

\begin{equation} M^{\alpha \beta } = \varepsilon ^{mnpq} S^{\alpha
}_{\ mn} S^{\beta }_{\ pq } \end{equation}

\noindent Then the matrix $M^{\alpha \beta }$ has a definite
signature (see (\ref G)). So, due to results of Urbantke \cite{U}
 and Harnett \cite{H}, the Urbantke metric (\ref I) is
non-degenerate, has a definite signature, and 2-forms $S^{\alpha }$
are self-dual or anti-self-dual with respect to Hodge operator
corresponding to this metric. Hence, the Urbantke metric can be
written as

\begin{equation} G_{mn} = \pm e^a_{\ m} e^a_{\ n}, \label L
\end{equation}

\noindent whereas 2-forms $S^{\alpha }$ as

\begin{equation} S^{\alpha } = C^{\alpha }_{\ \beta } {}^{-} \eta
^{\beta}_{\ ab} e^a \wedge e^b.  \label M \end{equation}

\noindent or

\begin{equation} S^{\alpha } = C^{\alpha }_{\ \beta } {}^{+} \eta
^{\beta }_{\ ab} e^a \wedge e^b, \label N \end{equation}

\noindent because the set of three 2-forms ${}^{\pm} \eta ^{\alpha
}_{\ ab} e^a \wedge e^b $ is a basis in the space of the
(anti)-self-dual forms.

One notes, that (\ref N) can be transformed in (\ref M). Indeed,
1-forms $e^a$ are defined by (\ref L) up to transformation

\begin{equation} e^a \longrightarrow O^a_{\ b} e^b, \ \ O \in O(4)
\label P \end{equation}

\noindent Let $O = \mbox{diag} \{ 1,-1,-1,-1 \}$ . Then

\begin{equation} {}^{+} \eta ^{\alpha }_{\ ab} O^a_c O^b_d = -
{}^{-} \eta ^{\alpha }_{\ cd } \label Q \end{equation}

\noindent So, redefining $e^a$ and $C^{\alpha }_{\ \beta }$
according to (\ref P) and (\ref Q), one can transform (\ref N) in
(\ref M).

One substitutes (\ref M) in (\ref G). Using the formulas of the
section 2.3, one obtains:

\begin{equation} C^{\alpha }_{\ \gamma }  C^{\beta }_{\ \gamma }=
\frac 1 3 \delta ^{\alpha \beta } C^{\delta }_{\ \gamma }  C^{\delta
}_{\ \gamma } \end{equation}

\noindent So

\begin{equation} C^{\alpha }_{\ \beta } = \pm C O^{\alpha }_{\ \beta
} \label R \end{equation}

\noindent where

\begin{equation} O \in SO(3), \ \  C = \sqrt {\frac 1 3 C^{\delta
}_{\ \gamma } C^{\delta }_{\ \gamma }} > 0.  \end{equation}

For given $O \in SO(3) $ there exists the matrix $\tilde O \in
SO(4)$ such that

\begin{equation} {}^{-}\eta ^{\alpha }_{\ ab} \tilde O ^a_{\ c}
\tilde O ^b_{\ d}= \left( O^{-1} \right)^{\alpha }_{\ \beta }
{}^{-}\eta ^{\beta }_{\ cd} \end{equation}

\noindent So, redefining $e^a $ according to (\ref P) with $O =
\tilde O$, and taking into account (\ref R), one reduces (\ref M) to

\begin{equation} S^{\alpha } = \pm C \, {}^{-}\eta ^{\alpha }_{\ ab}
e^a \wedge e^b \label S \end{equation}

Finally, substituting (\ref S) in (\ref I), one obtains that $C=1$.
The theorem is proved.

\subsection{$SU(3)$ YM action in Plebanski gauge}

We start from the usual $SU(3)$ YM action in the first order
formalism,

\begin{equation} S_{YM} = \int \mbox{tr} \left[ F \wedge (dA + A
\wedge A) + \lambda ^2 F \wedge *F \right], \label l \end{equation}

\noindent where $F$ and $A$ are considered as independent variables.

The forms $F$ and $A$ valued in the space of $3 \times 3$ anti-
Hermitian traceless matrices. So we can write

$$ A = \Gamma + i \Phi,$$

\begin{equation} F= S +iQ, \label m \end{equation}

\noindent where $\Gamma , \Phi, S, $ and $Q$ valued in the space of
real $3 \times 3$ matrices, and

$$ \Gamma ^T = - \Gamma , \ \ S^T = - S, $$

$$ \Phi ^T = \Phi, \ \ Q^T = Q, $$

\begin{equation} \mbox{tr} \Phi = 0, \ \ \mbox{tr} Q =0 \label n
\end{equation}

\noindent where the subscript $T$ means transposition.

Substituting (\ref m) in (\ref l), one obtains:

\begin{equation} S_{YM} = \int \mbox{tr}  \left[ S \wedge (R - \Phi
\wedge \Phi ) + Q \wedge D \Phi + \lambda ^2 S \wedge *S - \lambda
^2 Q \wedge *Q \right] \label p \end{equation}

\noindent where

\begin{equation} R = d \, \Gamma + \Gamma \wedge \Gamma ,
\end{equation}

\begin{equation} D \Phi = d \Phi + \Gamma \wedge \Phi + \Phi \wedge
\Gamma \label q \end{equation}

Decomposition (\ref m) corresponds to certain embedding of the
algebra $su(2) \approx o(3)$ in $su(3)$. So $\Gamma $ and $R$ can be
considered as the forms of connection and curvature corresponding to
the subgroup $SU(2)$ of the gauge group $SU(3)$ whereas $D$ is
covariant derivative defined by the connection $\Gamma $.

Due to (\ref n), one can write

$$ S^{\alpha \beta } = - \varepsilon ^{\alpha \beta \gamma}
S^{\gamma } $$

\noindent and to impose the gauge conditions (\ref G) on the 2-forms
$S^{\alpha }$.

Substituting (\ref K) in (\ref p), one obtains the action

$$ S = 2 \int e^A_{\ \, C'} \wedge e^{BC'} \wedge R^{AB} + 2 \int
e^{AC'} \wedge e^B_{\ C'} \wedge \Phi _A^{\ \, CDE} \wedge \Phi
_{BCDE} $$ $$+\int Q^{ABCD} \wedge D \Phi ^{ABCD} - \lambda ^2 \int
e^{AC'}\wedge e^B_{\ \, C'} \wedge *(e_A^{\ \, D'} \wedge e_{BD'} )
$$ \begin{equation} - \lambda ^2 \int Q^{ABCD} \wedge Q_{ABCD}
\label E \end{equation}

\noindent where $R_{AB}$ is defined by the formula (\ref{kk}),
 $\Phi^{ABCD}$ and $Q^{ABCD}$ are the forms $\Phi$, $Q$ written in
 the spinor language\footnote{We have wrote the YM action in new
 variables for the upper sign case in (\ref K). Below we will show
 that this is enough to formulate quantum version of the theory
 under consideration.}.

The first term in (\ref E) is the Palatini action. So the action
(\ref E) can be considered as one for gravity coupled with several
matter fields. In particular, first three terms in the action (\ref
E) are invariant under the action of the group of the general
coordinate transformations $Diff(R^4)$ - just as in general
relativity. But the total action, of course, is not $Diff(R^4)$
invariant because the last two terms in (\ref E) depend on fixed
space - time metric via Hodge operator.

We will continue the investigation of $SU(3)$ YM theory in
introduced variables in the next section at quantum level. But
before we must prove that Plebanski gauge really fixes the gauge up
to $SU(2)$ transformations.

\subsection{Investigation of the Plebanski gauge.}

Any $SU(3)$ matrix $U$ can be written in the form

\begin{equation} U = e^{i \omega } \end{equation}

\noindent where $\omega $ is a traceless Hermitian $3 \times 3$
matrix. Pure imaginary matrices $\omega $ corresponds to generators
of the subgroup $SU(2)$, whereas real matrices can be considered as
coordinates on the space $SU(3)/SU(2)$. Obviously, the latter
satisfy the equations

\begin{equation} \omega ^{\alpha \beta } = \omega ^{\beta \alpha},
\omega ^{\alpha \alpha } =0. \label t \end{equation}

Let us consider the infinitesimal gauge transformations with
parameters obeying (\ref t):

\begin{equation} \delta A = i d \omega + i [A, \omega ] \label {u'}
\end{equation}

\begin{equation} \delta F = i [F, \omega ] \label u \end{equation}

\noindent Comparing (\ref {u'}), (\ref {u}), and (\ref m), one
obtains:

\begin{equation} \delta \Gamma = - [\Phi , \omega ], \ \ \delta \Phi
= D \omega \end{equation} \begin{equation} \delta S = -[Q, \omega ],
\ \ \delta Q = [S, \omega ] \label x \end{equation}

Let

\begin{equation} T^{ABCD} = *\left[ S^{(AB} \wedge S^{CD)} \right]
\label {y'} \end{equation}

\noindent Using (\ref x) , one finds

\begin{equation} \delta T^{ABCD} = c \varepsilon^{mnpq}
e^{(A}_{\quad A'm} e^{B|A'|}_{\qquad \ n} Q^C_{\ \, EFGpq} \omega
^{D)EFG} \label y \end{equation}

\noindent where $c$ is irrelevant numerical constant,
$e^{AA'}_{\quad \, m}$ and $Q^{ABCD}_{\qquad \ pq}$ are components
of the forms $e^{AA'}$ and $Q^{ABCD}$, whereas  $\omega ^{ABCD}$ is
$SU(2)$ spinor that corresponds to $O(3)$ tensor $\omega ^{\alpha
\beta }$.

To prove that Plebanski gauge reduces the initial $SU(3)$ YM theory
to the $SU(2)$ one, it is sufficient to prove that the equations

\begin{equation} \delta  T^{ABCD} =0 \label z \end{equation}

\noindent have the only trivial solution $\omega ^{ABCD} = 0$ for
almost all field configurations.

One notes that due to (\ref t)

\begin{equation} \omega ^{ABCD} = \omega ^{(ABCD)}  \end{equation}

\noindent So eqs. (\ref z) are the system of five linear homogenous
equations for five unknown $\omega ^{ABCD}$.

The system (\ref z) can be rewritten as

\begin{equation} G^{(ABC}_{\quad \ \, EFG} \delta ^{D)}_H \omega
^{EFGH} =0 \label A \end{equation}

\noindent where

\begin{equation} G^{ABCEFG} = \varepsilon ^{mnpq} e^A_{\ \, A'm}
e^{BA'}_{\quad \, n} Q^{CEFG}_{\qquad pq} \label B \end{equation}

In generic case, the spinor $G^{ABCEFG}$ satisfies the only
constraints

$$G^{ABCDEF} = G^{(AB)CDEF}$$ \begin{equation} G^{ABCDEF} =
G^{AB(CDEF)} \end{equation}

Now let us consider the field configuration for which the only
non-zero components of $G^{ABCDEF}$ are $G^{000000}$, $G^{111111}$,
and

$$G^{000111}=G^{001011}=G^{001101}=G^{001110}. $$

\noindent For such configuration one obtains from (\ref A):

$$ G^{000000} \omega ^{0001} + G^{000111} \omega ^{1111} = 0 $$

$$ G^{000000} \omega ^{0000} = 0 , \ \ G^{000111} \omega ^{0011}=0,
$$

\begin{equation} G^{111111} \omega ^{0111} =0, \ \ G^{111111} \omega
^{1111} =0 \label C \end{equation}

Obviously, that the system (\ref C) has the only trivial solution
for non-zero $G^{000000}, G^{111111}$, and $G^{000111}$.

Let

\begin{equation} M^{ABCD}_{\qquad \, EFGH} = G^{(ABC}_{\quad \ \,
(EFG} \delta ^{D)}_{H)} \label D \end{equation}

\noindent the spinor $M^{ABCD}_{\qquad \, EFGH}$ can be considered {
as some $5 \times 5$ matrix $M$. We have proved that $\det M \not =
0$ for certain field configuration. But $\det M$ is polynomial with
respect to fields $e^{AA'}$ and $Q^{ABCD}$. So $\det M \not =  0$
for almost all field configuration. This means that the system (\ref
A) has the only trivial solution for almost all configurations of
fields.

\section{Quantization}

We start from usual expression for Euclidean vacuum expectation
value of certain Hermitian gauge invariant functional ${\cal O} =
{\cal O} [A, \Psi, \overline {\Psi}]$:

\begin{equation} < {\cal O} > = \int d A d \overline {\Psi} d \Psi
{\cal O} [A, \overline {\Psi} , \Psi ] \exp \{ -S_{YM} -S_{mat} \}
\label T \end{equation}

\noindent  where $\overline {\Psi} , \Psi $ are matter fields,

\begin{equation} S_{YM} [A] = - \frac{1}{4 \lambda  ^2} \int
\mbox{tr} (dA + A \wedge A) \wedge *(dA + A \wedge A),
\end{equation}

\noindent and

\begin{equation} S_{mat} = \int dx \sqrt g \left\{ \sum
\limits_{flavors} (\overline {\Psi} _f \widehat {\nabla} \Psi _f -
m_f \overline {\Psi} _f \Psi _f ) \right\} \end{equation}

Formula (\ref T) can be written as

$$ < {\cal O} > = \int d F d A d \overline {\Psi} d \Psi {\cal O}
[A, \overline {\Psi} , \Psi ] $$ \begin{equation}\exp \{ i \int
\mbox{tr} [F \wedge (dA + A \wedge A)] \} \exp \{ - \lambda ^2 \int
\mbox{tr} F \wedge *F - S_{mat} \} \end{equation}

\noindent or, finally, as

$$ < {\cal O} > = \int d S d Q d \, \Gamma d \Phi d {\overline \Psi}
d \Psi {\cal O}[\Gamma +i\Phi, \overline {\Psi} , \Psi ]$$
\begin{equation} \exp \{ i \int \mbox{tr} [S \wedge (R - \Phi \wedge
\Phi ) + Q \wedge D \Phi ] \} \exp \{ \lambda  ^2 \int \mbox{tr} [S
\wedge *S - Q \wedge *Q] \} \exp \{ -S_{mat} \} \label U
\end{equation}

\noindent where variables $S,Q, \Gamma ,$ and $ \Phi $ are defined
by (\ref m).

We will fix the gauge (namely, Plebanski gauge) by usual Faddeev -
Popov trick. We insert in (\ref U) the unit

\begin{equation} 1= \int \limits_{SU(3)/SU(2)}d \mu (\omega ) \delta
 \left( * [(S^{\omega })^{(AB} \wedge ( S^{\omega })^{CD)}] \right)
 \Delta _{FP} \end{equation}

\noindent where $d \mu (\omega )$ is invariant measure on
$SU(3)/SU(2)$ ,$(S^{\omega })^{AB} $is a gauge transformation of
$S^{AB}$, and $\Delta _{FP} $ is Faddeev - Popov functional. Then,
after usual manipulations, one obtains:

$$ < {\cal O} > = \int d S d Q d \, \Gamma d \Phi d \overline {\Psi}
d \Psi {\cal O}[\Gamma +i\Phi, \overline {\Psi} , \Psi ] \delta
\left( * [S^{(AB} \wedge  S^{CD)}] \right) \det M $$
 \begin{equation} \exp \{ i \int \mbox{tr} [S \wedge (R - \Phi
\wedge \Phi ) + Q \wedge D \Phi ] \} \exp \{ \lambda ^2 \int
\mbox{tr} [S \wedge *S - Q \wedge *Q] \} \exp \{ -S_{mat} \} \label
V \end{equation}

\noindent Here $\det M$ is Faddeev - Popov determinant, where $M$ is
$5 \times 5$ matrix

\begin{equation} M^{ABCD}_{\qquad \,  EFGH} =*\left[ S^{(AB} \wedge
Q^C_{\ \, (EFG} \delta ^{D)}_{H)} \right] \label w \end{equation}

\noindent This matrix coincides, on the surface

\begin{equation} S^{(AB} \wedge S^{CD)} =0, \label {w'}
\end{equation}

\noindent with the matrix (\ref D). So $\det M \not =0$ for almost
all field configurations (see the section 3.3).

Let ${\cal S} _{+} ({\cal S}_{-} )$ be the set of all 2-forms
$S^{\alpha }$ for which Urbantke metric (\ref I) is positive
(negative) definite.  We can write the integral (\ref V) as the sum
of the integrals over ${\cal S}_{+}$ and ${\cal S}_{-}$.

Obviously, that ${\cal S}_{-}$ is mapped onto  ${\cal S}_{+}$ by the
 transformation $S \longrightarrow -S, \ Q \longrightarrow -Q $.
 But the latter is equivalent to the complex conjugation in the
integral over ${\cal S}_-$.  So the integral over ${\cal S}_{+}$ is
equal to complex conjugated integral over $\cal S_{-}$.  Hence,

$$ < {\cal O} > = \mbox{Re} \int \limits_{{\cal S}_{+}} d S d Q d \,
\Gamma d \Phi d \overline {\Psi} d \Psi {\cal O}[\Gamma +i\Phi,
\overline {\Psi} , \Psi ] \delta \left( * [S^{(AB} \wedge S^{CD)}]
\right) \det M $$ \begin{equation}\exp \{ i \int \mbox{tr} [S \wedge
(R - \Phi \wedge \Phi ) + Q \wedge D \Phi ] \} \exp \{ \lambda  ^2
\int \mbox{tr} [S \wedge *S - Q \wedge *Q] \} \exp \{ -S_{mat} \}
\label X \end{equation}

We showed in the section 3.1 that the solution of Plebanski gauge
conditions (\ref {w'}) for $S \in {\cal S}_{+} $ is given, in vector
language, by the formula

\begin{equation} S^{\alpha } = {}^- \eta ^{\alpha }_{\  ab} e^a
\wedge e^b  \label Y \end{equation}

\noindent So, for $S \in {\cal S}_{+} $,

$$ \int \prod \limits_{a,n} de^a_n \prod \limits_{\alpha \atop {m >
n}} \delta (S^{\alpha }_{\  mn} - {}^- \eta ^{\alpha }_{\ mn} e^a_m
e^b_n) $$ \begin{equation} =f(S) \prod \limits _{\alpha \le \beta }
\delta ( *[S^{\alpha } \wedge S^{\beta } - \frac 1 3 \delta ^{\alpha
\beta } S^{\gamma } \wedge S^{\gamma }] ) \label Z \end{equation}

\noindent where the function $f(S)$ to be determined.

It is easy to prove that

\begin{equation} f=const \label {10} \end{equation}

\noindent Indeed,$f(S)$ is scalar density with respect to general
coordinate transformations and $O(3)$ gauge transformations. So

\begin{equation} f=f(\varepsilon ^{mnpq} S^{\alpha}_{\  mn}
S^{\alpha}_{\  pq}) \label {11} \end{equation}

\noindent But the dimension of $f$ is zero. So the function
(\ref{11}) is a constant.

Inserting (\ref{Z}) and (\ref{10}) in (\ref{X}), one obtains:

\begin{equation} < {\cal O} > = \mbox{Re} \int d e d Q d \, \Gamma d
\Phi d \overline {\Psi} d \Psi {\cal O}[\Gamma +i\Phi, \overline
{\Psi} , \Psi ] \det M \exp \{iS_1 - \lambda ^2 S_2 - S_{mat} \}
\label {12} \end{equation}

\noindent where

$$S_1 =\int e^A_{\ \, C'} \wedge e^{BC'} \wedge R_{AB}  + 2 \int
e^{AC'} \wedge e^B_{\  \, C'} \wedge \Phi _A^{\ \, CDE} \wedge
\Phi_{BCDE} + \int Q^{ABCD} \wedge D \Phi _{ABCD},$$
 \begin{equation}   S_2 = \int dx \sqrt{g} \, [ g^{mp}
 g^{nq}G_{mn}G_{pq} - (g^{mn}G_{mn})^2 ] +  \int Q^{ABCD} \wedge
*Q_{ABCD} \label {12a} \end{equation}

\noindent where

\begin{equation} G_{mn} = e^a_{\ m} e^a_{\  n} \label {12b}
\end{equation}

\noindent is YM induced metric, $\det M$ is Faddeev - Popov
determinant,

\begin{equation} M^{ABCD}_{\qquad \  EFGH} = *[e^{(A|C'|} \wedge
e^B_{\  C'} \wedge Q^C_{\  (EFG} \delta ^{D)}_{H)}] \end{equation}

\noindent and

$$ S_{mat} = \sum \limits_{flavors} \int dx \sqrt g \{ i \overline
{\Psi} _{fAB} \, {}^{\scriptscriptstyle (0)}e_a^{\  n} \gamma ^a D_n
\Psi _f^{\  AB} $$ $$ + i \overline {\Psi} _{fAB} \,
{}^{\scriptscriptstyle (0)}e_a^{\  n} \gamma ^a \Phi ^{ABCD}_{\ \  \
\  n} \Psi _{fCD} - m_f \overline {\Psi} _{fAB} \Psi _f^{\  AB} \}$$

\noindent In the latter formula ${}^{\scriptscriptstyle (0)}e_a^{\
n}$ means the space-time tetrad (that is, $g^{mn} =
{}^{\scriptscriptstyle (0)}e_a^{\  m} \, {}^{\scriptscriptstyle
(0)}e_a^{\  n }$).

The integrand in (\ref {12}) is $O(4)$ gauge invariant. To fix this
gauge freedom, it is necessary to impose further gauge conditions.
The simplest choice is

\begin{equation}  e_{ma} = e_{am} \label {13} \end{equation}

This gauge entangles space - time and gauge degrees of freedom and
so, after imposing of the gauge (\ref {13}), they must be considered
on the equal footing.

It is easy to prove, that Faddeev - Popov determinant, corresponding
to gauge (\ref {13}), is equal to $ |e|^{\frac 3 2}$. So the formula
(\ref{12}) can be written as

$$< {\cal O} > = \mbox{Re} \int \limits_{e_{am}=e_{ma}}d e d Q d \,
\Gamma d \Phi d \overline {\Psi} d \Psi {\cal O}[\Gamma +i\Phi,
\overline {\Psi} , \Psi ]  |e|^{\frac 3 2} \det M$$ \begin{equation}
\exp \{iS_1 - \lambda ^2 S_2 - S_{mat} \} \label {14} \end{equation}

\noindent (It would be remind, that $e^{AA'}_{\quad \  m}$ and
$e_{am}$ in (\ref {14}) are connected, according to rules of the
section 2.2, by relation $  e^{AA'}_{\quad \, m} =g^{aAA'} e_{am}).
$

The formula (\ref {12}) can be also rewritten in manifestly $O(4)$
invariant variables, such as $G_{mn}$,

$$ \Phi ^{mnpq}_{\ \ \ \ \ r} dx^r \equiv \Phi ^{\alpha \beta }_{\ \
r} \eta ^{\alpha}_{\  ab} e^a_{\  m}e^b_{\  n} \eta ^{\beta}_{\  cd}
e^c_{\ p} e^d_{\  q} dx^r $$

\noindent etc. But in such variables the corresponding action in
(\ref {12}) contains Einstein - Hilbert term $\sqrt G R$ and so is
not polynomial. So we prefer to consider formulas (\ref {12}) and
(\ref {14}) as final results of our investigation.

\section{Discussion.}

We have shown that gravity - like interactions live inside QCD. This
conclusion is supported by the results of Ne'eman and Sijacki
\cite{NS} concerning existence of gravity - like interactions in
infrared sector of QCD, and vice-a-versa.

Author hopes that the results presented in this paper will be
starting point of various new approaches to QCD. Here we will list
only some themes of the further investigations.

\begin{itemize} \item Rescaling fields $Q^{ABCD}$ and $G_{mn}$ in
(\ref {12}), one can rewrite this formula as

$$<{\cal O}>=$$ \begin{equation}  = \mbox{Re} \int  dedQd \, \Gamma
d \Phi d \bar \Psi d \Psi {\cal O}[\Gamma +i\Phi, \overline {\Psi} ,
\Psi ]  \det M \exp \{\frac{i}{\lambda} S_1 - S_2 - S_{mat} \}
\label {15} \end{equation}

So, at least in the weak coupling limit, it is naturally to
investigate the functional integral (\ref {15}) by stationary phase
method.

The stationary points are determined by equations

\begin{equation} \delta S_1 =0 \label {16}\end{equation}

But equations (\ref{16}) are nothing but Euclidean Einstein ones.
What is the meaning of known exact solutions of Einstein equations
(such as gravitational instantons, wormholes, etc.) in the context
of QCD?

\item In particular, what is the meaning of the flat space solution

\begin{equation} G_{mn} =c^2 g_{mn}, \ \ \ c=const \label {17}
\end{equation}

 \noindent of the equation (\ref {16})? How to construct the
expansion of the integrand in (\ref {15}) near such solution? Does
the existence of the flat solutions (\ref{17}) leads to appearance
the vacuum condensates of the gluon fields?

\item The actions $S_1,S_2, $ and $S_{mat}$ in (\ref{12}) are
polynomial. So it is possible to derive the corresponding Schwinger
equations. What are the solutions of these equations in the usual
approximations? Do the solutions exist that correspond to non-zero
vacuum condensate of the field $G_{mn}$?

\item The action $S_1 $ in (\ref{12}) is invariant with respect to
 the group $Diff(R^4)$ of the general coordinate transformations.
It is easy to derive the corresponding Ward identities. Obviously,
that these identities express nothing but the energy - momentum
conservation. Nevertheless, it is interesting to investigate
consequences of such Ward identities because in proposed variables
they have very unusual form and, most likely, can lead to new
interesting results.

\end{itemize}

Now let us discuss the shortcomings of the proposed approach. First,
our formulation is essencially chiral because left and right $SU(2)$
subgroups of the total $SU(2) \times SU(2) \approx O(4)$ invariance
group of the action (\ref E) play the different roles in our
formalism.  In itself, it is not a difficulty, but after imposing of
the gauge conditions that entangle space - time and internal degrees
of freedom (as the gauge (\ref{13})), one obtains the theory that is
not manifestly parity invariant. It is not convenient.

This left - right asymmetry in our approach is connected with the
structure of the group $SU(3)$. Indeed, there is  no faithful
embedding of the group $O(4)$ in $SU(3)$. So it is needed more large
gauge group to originate left - right symmetric general
relativity-like formalism. So

\begin{itemize} \item it is interesting to develop general
relativity-like formalism for the grand unified theories based on
the groups $SU(5),SO(10)$, etc. Except left - right symmetric
formulation, one may hope to find natural spontaneous parity
breaking mechanism in electroweak sector of the theory in this way.
\end{itemize}

Further, our theory is essential Euclidean and it is unclear how to
develop the general relativity-like formulation of YM theory in
which YM induced metric has Lorentzian signature in presented
approach. This shortcoming again is connected with the structure of
the gauge group $SU(3)$. Indeed, the gauge group $SU(3)$ is compact,
and so it is impossible to embed in $SU(3)$ neither the non-compact
group $SO(3,1)$ nor any its subgroup in a covariant way.

The existence of the only Euclidean formulation of the theory, per
se, is not a difficulty. But the formulation of the theory in the
Minkowski space is more visual. In particular, the absence of such
formulation hampers the investigation of the confinement in our
approach. Meanwhile, the results of the works \cite{L2,S} indicates
that, may be, there exists black hole like mechanism of the
confinement. But black holes live in the Lorentzian space rather
then in the Euclidean one.

\begin{itemize} \item Author hopes to overcome above mentioned
difficulties by using of formalism developed in the paper \cite{L4}
where it was shown that $SU(N)$ YM theory is equivalent to certain
$GL(N,C)$ gauge theory in the following sense: classes of the gauge
equivalent solutions of the initial $SU(N)$ YM theory are in
one-to-one correspondence to classes of the gauge equivalent
solutions of the above mentioned $GL(N,C)$ gauge theory. In the QCD
case $N=3$, and so Lorentz group $SL(2,C)$ can be embedded in the
QCD gauge group in such $GL(3,C)$ formalism. So it is possible to
develop the Lorentzian analog of the Euclidean general relativity
like formulation of QCD given in the presented work.  \end{itemize}


\begin{thebibliography}{99}

\bibitem{HNH} F. W. Hehl, J.Nitch, and P. von der Heyde, in {\em
General Relativity and Gravitation -- One Hundred Years After the
Birth of Albert Einstein,} p.329 (Plenum, New York,1980)

\bibitem{M1} E. W. Mielke, {\em Geometrodynamics of Gauge Fields}
(Academic - Verlag, Berlin,1987)

\bibitem{SIS} A. Salam, ICTP preprint IC/71/3 (unpublished);
C.~J.~Isham, A.~Salam, and J.~Strathdee, Phys. Rev. {\bf D8}, 2600
(1973); C.~Sivaram and K.~P.~Singha, Phys. Rep. {\bf 51}, 111
(1979); A.~Salam and C.~Sivaram, Mod. Phys. Lett., {\bf A8}, 321
(1993)

\bibitem{M2} E. W. Mielke, Int. J. Theor. Phys., {\bf 19}, 189
(1980)

\bibitem{HD} J. C. Huang and P. W. Dennis, Phys. Rev. {\bf D21}, 910
(1980); {\bf D23}, 1723 (1981); {\bf D24}, 35 (1981)

\bibitem{ISS} C. J. Isham, A. Salam, and J.~Strathdee, Phys. Rev.
{\bf D3}, 867 (1971)

\bibitem{L1} F. A. Lunev, Phys. Lett. {\bf B295}, 99 (1992); Theor.
Math. Phys., {\bf 94}, 66 (1993)

\bibitem{J} K. Johnson, in {\em QCD -- 20 Years Later}, p.795 (World
Scientific, 1993)

\bibitem{L2} F. A. Lunev, Phys. Lett. {\bf B311}, 273 (1993)

\bibitem{L3} F. A. Lunev, Phys. At. Nucl., {\bf 56}, 1591 (1993)
(Russian edition: Yad. Fiz., {\bf 56}, 238 (1993))

\bibitem{HOM} F. W. Hehl, E. W. Mielke, and Yu. Obuhov, Phys. Lett.
{\bf A192}, 153 (1994)

\bibitem{BFH} M. Bauer, D. Z. Freedman, and P.~E.~Haagensen, Nucl.
Phys. {\bf B} (to be published)

\bibitem{HJ} P. E. Haagensen, K. Johnson, Nucl. Phys. {\bf B} (to be
published), hep-th/9408164

\bibitem{RS} V. Radovanovic and D. Sijacki (unpublished),
hep-th/9411002

\bibitem{FHJL} D. Z. Freedman, P. E. Haagensen, K.~Johnson, and
J.-I. Lattore, preprint CERN-TH 7010/93 (unpublished)

\bibitem{S} D. Singlton (unpublished), hep-th/9501052,950197,9502116

\bibitem{SS} A. Salam and  J. Strathdee, Phys. Rev., {\bf D16}, 2668
(1977)

\bibitem{P} J. F. Plebanski, J. Math. Phys., {\bf 18}, 2511 (1977)

\bibitem{JS} T.Jacobson, L. Smolin, Class. Quant. Grav., {\bf 5},
583 (1988)

\bibitem{Sam} J. Samual, Pranama, {\bf 28}, L429 (1987)

\bibitem{NS} Y. Ne'eman and Dj. Sijacki, Phys. Lett., {\bf B247},
571 (1990); {\bf B276}, 173 (1992); preprint CERN-TH. 7461/94
(unpublished); preprint TAUP N232-94 (unpublished)

\bibitem{K} M. Yu. Kuchiev, Europhys. Lett., {\bf 28}, 539 (1994)

\bibitem{Pel} P. Peldan, Nucl. Phys., {\bf B395}, 239 (1993)

\bibitem{U} H. Urbantke, J. Math. Phys., {\bf 25}, 2321 (1984)

\bibitem{H} G. Harnett, J. Math. Phys., {\bf 32}, 84 (1991); J.
Phys., {\bf A25}, 5649 (1992)

\bibitem{B} I. Bengtsson, preprint USITP 95-2 (unpublished),
gr-qc/9502010

\bibitem{'tH} G. 't Hooft, Nucl. Phys., {\bf B357}, 211 (1991)

\bibitem{A} A. Ashtekar, Phys. Rev. Lett. {\bf 57}, 2244 (1986);
Phys. Rev., {\bf D36}, 1587 (1987)

\bibitem{CDJM} R. Capovilla, J. Dell, T.~Jacobson, and L.~Mason,
Class. Quant. Grav., {\bf 8}, 41 (1991)

\bibitem{Hal} M. B. Halpern, Phys. Rev., {\bf D16}, 3515 (1977)

\bibitem{L4} F. A. Lunev, Mod. Phys. Lett., {\bf A9}, 2281 (1994)

\bibitem{CP} S. Chakraborty and P. Peldan, preprints CGPG - 94/1-3,
CGPG - 94/2-4 (unpublished) gr-qc/9401028, 9403002

\bibitem{Rob} D. C. Robinson, Class. Quant. Grav., {\bf 11}, L157
(1994); King's Colledge preprint (1995) (unpublished)

\end{thebibliography}
\end{document}